\documentclass[sigchi-a]{acmart}

\AtBeginDocument{%
  \providecommand\BibTeX{{%
    \normalfont B\kern-0.5em{\scshape i\kern-0.25em b}\kern-0.8em\TeX}}}

\PassOptionsToPackage{hyphens}{url}\usepackage{hyperref}

\usepackage{color, colortbl}
\definecolor{Gray}{gray}{0.95}
\definecolor{LightCyan}{rgb}{0.88,1,1}
\definecolor{Yellow}{RGB}{252, 243, 207}
\setcopyright{rightsretained}

\begin{document}

\copyrightyear{2019} 
\acmYear{2019} 
\acmConference[CSCW '19 Companion]{2019 Computer Supported Cooperative Work and Social Computing Companion Publication}{November 9--13, 2019}{Austin, TX, USA}
\acmBooktitle{2019 Computer Supported Cooperative Work and Social Computing Companion Publication (CSCW '19 Companion), November 9--13, 2019, Austin, TX, USA}
\acmDOI{10.1145/3311957.3359501}
\acmISBN{978-1-4503-6692-2/19/11}

%\title[Poster Presentation]{Revealing Differences in Semantic Context on a Data Privacy Online Discussion: An Inter-Language Social Media Study}
\title[Poster Presentation]{Information Privacy Opinions on Twitter: A Cross-Language Study}

\author{Felipe Gonz\'{a}lez}
\affiliation{%
  \institution{UTFSM}
  \city{Santiago}
  \country{Chile}
}
\email{felipe.gonzalezpi@usm.cl}

\author{Andrea Figueroa}
\affiliation{%
  \institution{UTFSM}
  \city{Valpara\'{i}so}
  \country{Chile}
}
\email{andrea.figueroa@usm.cl}

\author{Claudia L\'{o}pez}
\affiliation{%
  \institution{UTFSM}
  \city{Valpara\'{i}so}
  \country{Chile}
}
\email{claudia@inf.utfsm.cl} 

%\author{Yihan Yu}
%\affiliation{%
  %\institution{University of Washington}
  %\city{Seattle}
  %\country{United States}
%}
%\email{yyu2016@uw.edu}

\author{Cecilia Aragon}
\affiliation{%
  \institution{University of Washington}
  \city{Seattle}
  \country{United States}
}
\email{aragon@uw.edu}

\renewcommand{\shortauthors}{González, et al.}

\begin{abstract}

The Cambridge Analytica scandal triggered a conversation on Twitter about data practices and their implications.
Our research proposes to leverage this conversation to extend the understanding of how information privacy is framed by users worldwide. 
{\let\thefootnote\relax\footnote{{This collaboration was possible thanks to the support of the Fulbright Program, under a 2017-18 Fulbright Fellowship award. This work was also partially funded by CONICYT Chile, under grant Conicyt-Fondecyt Iniciaci\'on 11161026. 
The first author acknowledges the support of the PIIC program from Universidad T\'{e}cnica Federico Santa Mar\'{i}a and CONICYT-PFCHA/Mag\'{i}sterNacional/2019-22190332}}}
We collected tweets about the scandal written in Spanish and English between April and July 2018. We created a word embedding to create a reduced multi-dimensional representation of the tweets in each language.  
For each embedding, we conducted open coding to characterize the semantic contexts of key concepts: ``information'', ``privacy'', ``company'' and ``users'' (and their Spanish translations). Through a comparative analysis, we found a broader emphasis on privacy-related words associated with companies in English. We also identified more terms related to data collection in English and fewer associated with security mechanisms, control, and risks. Our findings hint at the potential of cross-language comparisons of text to extend the understanding of worldwide differences in information privacy perspectives.

\end{abstract}

\maketitle

\section{INTRODUCTION}
\begin{margintable}
\vspace{10mm}
\caption{Datasets before and after data cleaning}
\label{Table:NumberOfTweets}
%\resizebox{\columnwidth}{!}{%
\begin{tabular}{@{\extracolsep{6pt}}llrrrr}
\toprule
Dataset             & \multicolumn{2}{c}{Spanish}  & \multicolumn{2}{c}{English}              \\ 
    \cmidrule{2-3} 
    \cmidrule{4-5} 
                    & \#Tweets      & \#Users       & \#Tweets      & \#Users   \\
\midrule
Total               & 472,363     & 222,352     & 7,476,988   & 1,846,542   \\
Original    & 106,656     & 47,951      & 1,572,371   & 574,452       \\
%Analyzed         & 100,606 & 44,182 &  1,442,112 &  504,214\\ 
\textbf{Human}   &           \textbf{74,644} & \textbf{36,056}      & \textbf{975,678} & \textbf{410,180}\\
%\textbf{Humans}   &           \textbf{74,644} & \textbf{36,056}      & \textbf{983,514} & \textbf{414,161}      \\ 

\bottomrule
\end{tabular}
%}
\end{margintable}
%``Information privacy refers to the desire of individuals to control or have some influence over data about themselves'' \cite{belanger2011privacy}. 
Information privacy % in Information Systems research have frequently 
has been defined as ``the ability of individuals to control the terms under which their personal information is acquired and used'' \cite{culnan2003consumer}. According to public opinion polls, privacy is one of the major concerns of people nowadays \cite{smith2011information}. %In particular, information privacy has long been a concern for users of social network sites \cite{li2018sns}. 
Different measurement instruments have been developed to identify, analyze and evaluate privacy concerns \cite{yun2019chronological, belanger2011privacy}. Two well-known instruments are the Concern for Information Privacy (CFIP) \cite{smith1996information} and Internet Users' Information Privacy Concerns (IUIPC) \cite{malhotra2004internet}. IUIPC adapts CFIP into the internet context \cite{smith2011information} and is widely used even today \cite{yun2019chronological, raber2018privacy}. %The latter adapts CFIP into the internet context \cite{smith2011information} and it is widely used until today \cite{yun2019chronological, raber2018privacy}.

While it is expected that individuals from different world regions have different cultures, values and laws that can result in differences in their perceptions of information privacy and its impacts \cite{belanger2011privacy}, there is still a limited understanding of such differences. % on worldwide perceptions of privacy concerns are still 

%The Cambridge Analytica scandal unfolded that 

In 2018, it was revealed that the personal data of 87 million Facebook users were exposed and used by Cambridge Analytica to support political campaigns \cite{lapaire2018content}. This Cambridge Analytica scandal sparked a worldwide conversation on Twitter about this particular misuse of user data. Our project seeks to use these online public communications %considering the IUIPC instrument 
to identify differences and similarities on data privacy perspectives by people who write in different languages, which we see as a proxy to represent different world regions. We report our preliminary results and discuss their potential to deepen the understanding of information privacy perspectives worldwide.  %from different speaking languages.
%Prior privacy research has argued that language and country of residence can relate to diverse perspectives on privacy \cite{gonzalez2019global}.

\section{Related Work and research questions} %definir IUIPC, diferencias culturales. 

%The goal of t
The IUIPC instrument was designed to reflect and identify Internet users' concerns about information privacy from a user perspective of fairness %considering fairness/justice individuals' perception 
\cite{malhotra2004internet}. It contains three dimensions of concerns: the trade-off between personal data \textit{collection} by others and perceived benefits, the users' ability to \textit{control} their personal information, and their \textit{awareness} of organizational information privacy practices.
%\begin{itemize}
    %\item Control: Individual concerns about control over personal information (eg. approval, modification, opt-out).
    %\item Awarenes : Individual concerns about her/his awareness of organizational information privacy practices.
    %\item Collection: Individuals' concerns about the amount of personal data possessed by others relative to the value of benefits received. 
%\end{itemize}
%%Indicar ejemplos de donde se ha utilizado IUIPC

Most research on privacy concerns has been conducted through questionnaires \cite{ur2013cross}, such as the IUIPC \cite{raber2018privacy}. While widely used to understand personal privacy concerns in North America and Europe (e.g., \cite{kaya2003cross, dinev2006privacy}), these surveys have been less frequently applied in other world regions. Thus, this tendency has left open many research questions about how privacy perspectives vary across the globe. 

Recent work has explored text mining as an alternative research method. Raber and Kr{\"u}ger found that IUIPC dimensions can be derived from written text \cite{raber2018privacy}. They observed a correlation between IUIPC concerns, as measured by a questionnaire,
and LIWC language features of social media posts from 100 users. Inspired by this line of work, we propose to analyze the semantic context of privacy-related words in online communications in different languages to explore its potential for revealing worldwide differences in information privacy perspectives. In particular, we seek to use tweets about the Cambridge Analytica scandal in two languages as a corpus to observe similarities and differences in how people conceptualize information privacy.    
%The context in which a discussion is held can influence how arguments are understood \cite{rho2018fostering} and reveal individual concerns. Considering that the way people communicate about a social online discussion could influence their perception of related topics \cite{rho2018fostering} and that people from different language may discuss the same issue from a different point of view, we 
%%%%% 
Two research questions guide our work: %\textbf{RQ1:} Do IUIPC dimensions emerge from online communications after a data breach scandal?, \textbf{RQ2:} Are there differences in the semantic contexts of privacy-related words that come from online communications written in two different languages? 
\begin{itemize}
   \item \textbf{RQ1:} Do IUIPC dimensions emerge from online communications after a data breach scandal?
    \item \textbf{RQ2:} Are there differences in the semantic contexts of privacy-related words that come from online communications written in two different languages? 
\end{itemize}

%This implies that the way people talk about online social movements could in fact, influence their perception of related topics (cscw 2018 paper)

%Word embeddings can reveal latent contextual cues of tokens by capturing the co-occurence of terms with an associated word\cite{rho2018fostering}.

%Given the relative lack of data privacy research in international settings \cite{smith2011information}

\section{Data and Methods}
\begin{margintable}
\vspace*{-40mm} 
\centering
\caption{Emergent categories during open coding}
\label{tab:categories_open_coding}
\begin{tabular}{p{0.38\linewidth}p{0.88\linewidth}} %la suma tiene que hacer 95%. 
\toprule
Categories & Description \\ \bottomrule
\rowcolor{Yellow}
Data \& Information & Direct references to these concepts and examples of user data and information%Information and material for the generation of information 
\\
\hline
\rowcolor{Yellow}
Companies \& Organizations & Entities that manipulate user data for their own purposes \\
\hline
\rowcolor{Yellow}
Users & Data owners%Users as owners of data and/or information 
\\
\hline
\rowcolor{LightCyan}
Data collection, handling and/or storage & %Technology or technique to obtain, collect and/or handle data 
Technologies or techniques to obtain, collect and/or handle data\\
\hline
\rowcolor{LightCyan}
Privacy \& security terms & Words associated with data privacy and security %of data and information 
\\
\hline
\rowcolor{LightCyan}
Security mechanisms & Tools and techniques that %are used to 
implement security services \\
\hline
\rowcolor{LightCyan}
Privacy \& security risks & Entities or bad practices that can compromise sensitive data \\
\hline
\rowcolor{LightCyan}
Ownership agency & %Concerns about 
Control over personal information\\
\hline
\rowcolor{LightCyan}
Regulation & Law, rule or regulation that controls the use of user data %and/or information of users 
\\
\hline
\rowcolor{Gray}
Synonymous & %The nearby word and the token 
Words with the same meaning than the token\\
\hline
\rowcolor{Gray}
Attribute or characteristic & %The nearby term represents 
A characteristic of the token \\
\hline
\rowcolor{Gray}
Action & Action or activity linked to the token\\
\hline
\rowcolor{Gray}
Third party & Entity that can not be categorized as User or Company because there is not sufficient contextual information to do so \\
\hline
\rowcolor{Gray}
Reaction or attitude & %Response to a stimulus. 
Way of feeling or acting toward a person, thing or situation. \\
\hline
\rowcolor{Gray}
Undetermined & The relation between the token and the %nearby 
word is not exactly known, established or defined \\
\bottomrule
\end{tabular}
\centering
\scriptsize{\textbf{Token-category:} yellow \textbf{Privacy:} light blue  \textbf{Other:} gray}
\end{margintable}
To answer our research questions, we collected tweets written in Spanish and English between April 1st and July 10th, 2018. We used the Twitter API to capture tweets that include hashtags or keywords related to the Cambridge Analytica scandal or data privacy, such as ``\#CambridgeAnalytica'' %, ``\#DeleteFacebook'', ``Zuckerberg''
and ``Facebook privacy''. We retrieved more than 7.4 million tweets written in English, and more than 470,000 tweets in Spanish (see Table \ref{Table:NumberOfTweets}). %The tweets in English were generated by about 1.8 million unique Twitter accounts, while tweets in Spanish were produced by approximately 220,000 users.%To answer our research questions, we used Tweepy%\footnote{http://www.tweepy.org/} , a Python library for accessing to the standard realtime streaming Twitter API. Using this library we were able  to capture tweets that include hashtags or keywords related to the Cambridge Analytica scandal or data privacy, such as: ``\#CambridgeAnalytica'', ``\#DeleteFacebook'', ``Zuckerberg'' and ``Facebook privacy''. The  standard realtime streaming Twitter API returns
%a random sample of all public tweets that match the search keywords. We collected tweets written in Spanish and English between April 1st and July 10th, 2018. Overall, we collected more than 7.4 million tweets written in English and more than $470,000$ tweets in Spanish (see Table \ref{Table:NumberOfTweets}). The English tweets were generated by about 1.8 million unique Twitter accounts while the Spanish tweets were produced by approximately 220,000 users. The difference between the number of tweets and users collected in English and Spanish suggests that English-speaking Twitter users tweeted more about this scandal using the selected keywords than Spanish-speaking users, although this may be explained by the greater volume of English tweets overall%\footnote{https://www.statista.com/statistics/267129/most-used-languages-on-twitter/}.

We cleaned our dataset in two ways. First, we removed all retweets to focus on original opinions.  %and avoid analyzing duplicates. 
This step downsized both datasets by 80\%. %to approximately 20\% of their original sizes. 
Second, we attempted to eliminate tweets generated by automated accounts so our study could indeed reflect people's opinions. %To do so, 
We chose Botometer \cite{davis2016botornot} to identify potential bots. More details about this process can be found in \cite{gonzalez2019global}. Our final dataset includes %36,056 users who tweeted 
74,644 tweets in Spanish and %about the Cambridge Analytica scandal. 
%The English dataset comprises 410,180 accounts that generated 
975,678 tweets in English (see Table \ref{Table:NumberOfTweets}). % details these figures. More details about this process can be found in \cite{gonzalez2019global}.

Following Rho et al.'s approach \cite{rho2018fostering}, we used word embeddings \cite{mikolov2013distributed} to analyze the semantic context in which a concept under study is framed. Based on co-occurrence of terms, word embeddings create a reduced multi-dimensional representation of a corpus of text that allows assessing the semantic proximity among terms in a corpus. Thus, analyzing the closest terms of a given word can reveal the context in which it is used \cite{rho2018fostering}.

%Word embedding is one of the most popular representation of document vocabulary. It is capable of capturing context of a word in a document, semantic and syntactic similarity, relation with other words, etc.

To enable a cross-language comparison, we built a word embedding for all tweets written in the same language. Before creating them, we removed stopwords and transformed the text to lowercase. We customized our stopwords to ensure that digits and symbols like ``\#'' were removed but not the words that contain it. Links and usernames were removed. As a result, our corpus comprised 76,128 unique words in English and 21,736 in Spanish.

%Linguistic analysis is also a useful method to evaluate how people react to a topic based on their commenting behavior \cite{rho2018fostering}. Creating a word embedding on Facebook comments for three politically distinct news publishers and analyzing nearby words to the token ``Metoo'' in each of these, it was possible to identify differences about the context in which the movement \#Metoo was discussed by people follower of different political news publishers \cite{rho2018fostering}.
%Understanding how a word is characterized by its nearby words can reveal key linguistic contexts in which it is discussed \cite{rho2018fostering}. 

%the tokens: ``information'', ``privacy'', ``users'' and ``company'' and nearby words that co-occur with it. Before vectorizing our words, 

%To choose the best word embedding architecture, 
We considered seven word embedding architectures that involve \textit{Word2Vec/FastText}, \textit{CBOW/Skipgram}, and different numbers of dimensions and epochs. Each word embedding architecture for the English corpus was evaluated over 15 %intrinsic conscious word embeddings 
evaluation methods \cite{jastrzebski2017evaluate} (e.g., \textit{Google Analogy Test Set}, \textit{MTurk-287}, \textit{ESSLI\_1a}). % using a word embedding benchmark library.\footnote{https://github.com/kudkudak/word-embeddings-benchmarks}
Considering all terms that appear at least 3 times %in the corpus 
and a %conservative 
window of 5 terms, a Word2Vec CBOW architecture with 300 dimensions trained during 50 epochs achieved the best performance. The same architecture was used for the Spanish corpus. Gender bias in our embeddings was reduced using Bolukbasi's methodology \cite{bolukbasi2016man}. 

To analyze the semantic
context in which information privacy was framed in Spanish and English, two of the authors conducted open coding of the 40 closest words to four privacy-related tokens: \textit{privacy}, \textit{information}, %. Also, to reveal privacy concerns about users and organizations data practices the tokens: 
\textit{users}, and \textit{company}. %were used as well. In the case of the Spanish word embedding, 
Their respective Spanish translation were also used: \textit{privacidad}, \textit{informaci\'{o}n}, \textit{usuarios}, and \textit{empresa}. %Content analysis data is available at \url{https://github.com/gonzalezf/Information-Privacy-Opinions-on-Twitter-A-Cross-Language-Study}. 
Through an iterative process, the coders consolidated the open codes into 15 categories (see Table \ref{tab:categories_open_coding}). A total of 320 words were independently re-classified in these categories. Considering the eight tokens, the average Cohen's kappa score was 0.707.

%(range: [0.630,0.797]).
%The Kappa scores of the categorization %0.707 in average ( min =0.630, max = 0.797)
%of 0.797 for nearby words to ``information'', 0.685 for ``privacy'', 0.717 for ``'users'' and 0.630 for ``company''.

%, users and company (and their repectively Spanish translations: \textit{privacidad}, \textit{informaci\'{o}n}, \textit{usuarios} y \textit{empresa} in the English and Spanish word embedding respectively. 

\section{Results and conclusions}
\begin{margintable}
\vspace*{-40mm} 
\center
\caption{Representation of privacy categories per each token in the Spanish and English word embeddings }
\label{tab:privacy_representation}
%\resizebox{\columnwidth}{!}{%
\begin{tabular}{@{\extracolsep{6pt}}lrr}
\toprule
& Spanish (\%)  & English (\%)  \\ 
\midrule
Company &  2.50 & 12.50 \\
Users &  30.00 &  30.00 \\
Information & 32.50 &  27.50 \\
Privacy &  62.50 & 62.50\\
\bottomrule
\end{tabular}
%}
\end{margintable}

\begin{marginfigure}
\vspace{20mm}
\centering
\includegraphics[width=1.3\linewidth]{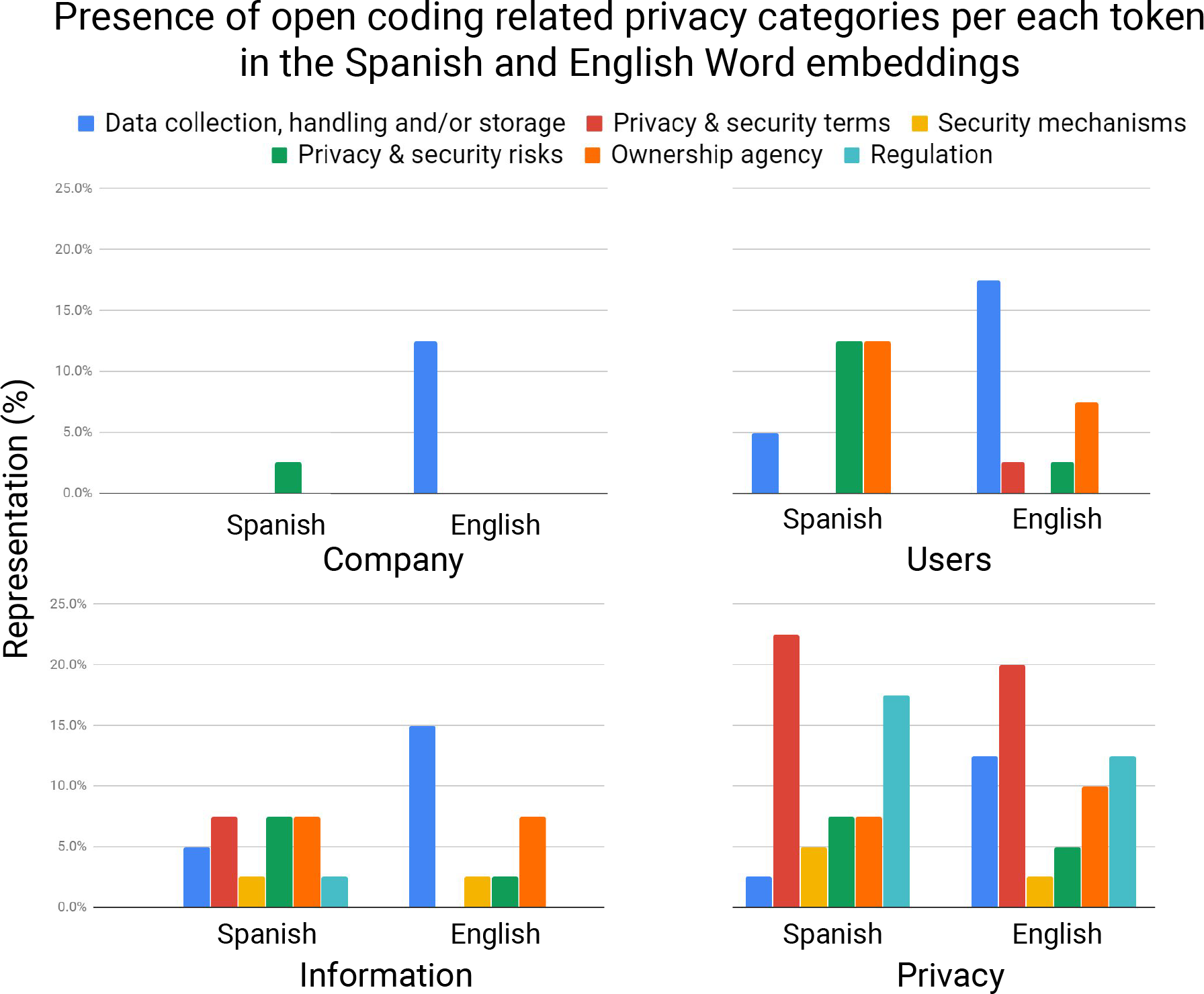}
\caption{Representation of data privacy categories in each word embedding }
\label{fig:iuipc_dimensions_per_language_details}
\end{marginfigure}

%We developed a taxonomy  using grounded theory techniques \cite{strauss1990grounded} in order to describe the nature of nearby terms to information privacy such as: ``information'' and  ``privacy''. Also,  to identify concerns about organizations and users data practices we

%During the open coding phase, we uncovered 15 different categories (see Table \ref{tab:categories_open_coding}). 

The coding process resulted in 15 categories. Three of them match our initial tokens: \textit{information}, \textit{company}, and \textit{users}. %can be directly associated with one open coding category. For example, the token company (``\textit{empresa}'' in Spanish) is represented with the Companies \& Organizations category). 
Instead, the token \textit{privacy} can be associated with six different categories. Answering RQ1, five of them are related to the IUIPC dimensions.
%Some of the categories found can be associated with the  collection, control and awareness IUIPC dimensions. 

The IUIPC's \textit{collection} dimension refers to the ``degree to which a person is concerned about the amount of individual-specific data possessed by others relative to the value of benefits received'' \cite{malhotra2004internet}. Through open coding, we identify a \textbf{data collection, handling and/or storage} category that contains words associated with technology or techniques useful to obtain, collect or handle data, including databases, services, app and website. The \textit{control} dimension denotes concerns about control over personal information. This is often exercised through approval, modification and opportunity to opt-in or opt-out \cite{malhotra2004internet}. Terms related to this dimension appear in the coding phase (e.g., consent, opt, permission) and are categorized as \textbf{ownership agency}. This category also includes %words associated with 
advice directed to users and good privacy practices %directed to the users 
%were also grouped in this category 
(e.g., %\textit{prevenir} (
prevent%), \textit{protege} (
, protect %), \textit{evitar} (
, and avoid in Spanish along with \textit{cuidatusdatos}, which means take care of your data). %In our model, we call this category as \textbf{ownership agency}. 
The third IUIPC dimension is \textit{awareness} that % of privacy practices, can be understood as the 
refers to individual concerns about her/his awareness of organizational information privacy practices \cite{malhotra2004internet}. Three of our categories are associated with this dimension. %First, the category 
\textbf{Privacy and security terms} comprise words such as %contains words associated with the privacy and security of data and information (e.g 
confidentiality, transparency, safety, \textit{seguridaddigital} (digital security), \textit{ciberseguridad} (cybersecurity). %Second, the category 
\textbf{Security mechanisms} %describe tools and techniques that are used to implement security services  (e.g: 
include, among others, the following terms: \textit{contrase\~nas} (passwords) and encryption. Finally, %The third and last category, 
\textbf{privacy \& security risks} %is identified with 
refers to entities or bad practices that can compromise sensitive data, for example: \textit{troyano} (trojan), databreach, grooming, \textit{ciberdelincuente} (cybercriminal).

Through a comparative analysis of categories by token, we observe that a broader proportion of words is covered by privacy-related categories in English than in Spanish (see Table \ref{tab:privacy_representation}). % is bigger in the English word embedding %than Spanish word embedding. 
%than the Spanish word embedding. For example, 
This difference is largely explained by the context around the \textit{company} token, which is 5 times larger in English. %This is related with previous work, were it was found that English Twitter users tend to discuss the responsibilities of companies in the Cambridge Analytica scandal \cite{gonzalez2019global}. 

% That is not the case of the English word embedding, where only when the nearby words to privacy were labeled the awareness dimension represent the majority of tokens. In the English word embedding, the discussion is focus in the data collection, handling or storage (see Figure \ref{fig:iuipc_dimensions_per_language_details}). 

Regarding the %to the representation of each 
IUIPC dimensions, there is a broader emphasis on \textit{collection} as the category
%per each token,  the discussion around 
data collection, handling and/or storage %is more representative 
cover more words in the English tokens, compared to the Spanish ones (see Figure \ref{fig:iuipc_dimensions_per_language_details}). %English Twitter users are more focused on the collection IUIPC dimension than Spanish Twitter users.
In turn, 
we note %identified that per each token 
that the three categories related to %the presence of the 
\textit{awareness} (privacy and security terms, risks and security mechanisms) cover more words in Spanish than English. %In Spanish there is a tendency to discuss the event focusing in the privacy and security risks, security mechanism and regulations. On the other hand,  
Lastly, %for both languages, the discussion about 
ownership agency (\textit{control} in IUIPC) appears slightly more in Spanish, especially regarding the token \textit{user}. %rates in both languages. %emerges around the tokens: ``users'', ``information'' and ``privacy''. %These findings reflects Spanish and English Twitter users' concerns in the context of a online data privacy breach discussion. 

{\let\thefootnote\relax\footnote{{Full data is available at \url{https://github.com/gonzalezf/Information-Privacy-Opinions-on-Twitter-A-Cross-Language-Study}}}}Finally, another privacy-related category that emerges from our coding but can not be directly associated with the IUIPC dimensions is \textbf{regulations}. This category was highly relevant when analyzing the semantic context of the token \textit{privacy} in both languages, but slightly more in Spanish. 

Overall, the preliminary results reported here suggest that social media text written in two languages can be used to reveal different emphasis on information privacy perspectives (RQ2). %The categories we identified can be related to current privacy models, but also have the potential to extend our understanding of privacy through the emergence of new concepts. %, such as the case of regulations and 
\bibliographystyle{ACM-Reference-Format}
\bibliography{references}

\end{document}